# Tightening LP Relaxations for MAP using Message Passing

**David Sontag**
CSAIL, MIT
Cambridge, MA

**Talya Meltzer**
Hebrew University
Jerusalem, Israel

**Amir Globerson**
CSAIL, MIT
Cambridge, MA

**Tommi Jaakkola**
CSAIL, MIT
Cambridge, MA

**Yair Weiss**
Hebrew University
Jerusalem, Israel

## Abstract

Linear Programming (LP) relaxations have become powerful tools for finding the most probable (MAP) configuration in graphical models. These relaxations can be solved efficiently using message-passing algorithms such as belief propagation and, when the relaxation is tight, provably find the MAP configuration. The standard LP relaxation is not tight enough in many real-world problems, however, and this has lead to the use of higher order *cluster-based* LP relaxations. The computational cost increases exponentially with the size of the clusters and limits the number and type of clusters we can use. We propose to solve the cluster selection problem monotonically in the dual LP, iteratively selecting clusters with guaranteed improvement, and quickly re-solving with the added clusters by reusing the existing solution. Our dual message-passing algorithm finds the MAP configuration in protein side-chain placement, protein design, and stereo problems, in cases where the standard LP relaxation fails.

## 1 Introduction

The task of finding the maximum aposteriori assignment (or MAP) in a graphical model comes up in a wide range of applications. For an arbitrary graph, this problem is known to be NP hard [11] and various approximation algorithms have been proposed.

*Linear Programming (LP) relaxations* are commonly used to solve combinatorial optimization problems in computer science, and have a long history of being used to approximate the MAP problem in general graphical models (e.g., see [9]). LP relaxations have an advantage over other approximate inference schemes in that they come with an optimality guarantee – if the solution to the linear program is integral, then it is guaranteed to give the global optimum of the MAP problem.

An additional attractive quality of LP relaxations is that they can be solved efficiently using message-passing algorithms such as belief propagation and its generalizations [3, 13, 15]. In particular, by using message-passing algorithms, we can now use LP relaxations for large-scale problems where standard, off-the-shelf LP solvers could not be used [18].

Despite the success of LP relaxations, there are many real-world problems for which the basic LP relaxation is of limited utility in solving the MAP problem. For example, in a database of 97 protein design problems studied in [18], the standard LP relaxation allowed finding the MAP in only 2 cases.

One way to obtain tighter relaxations is to use *cluster-based* LP relaxations, where local consistency is enforced between cluster marginals. As the size of the clusters grow, this leads to tighter and tighter relaxations. Furthermore, message-passing algorithms can still be used to solve these cluster-based relaxations, with messages now being sent between clusters and not individual nodes. Unfortunately, the computational cost increases exponentially with the size of the clusters, and for many real-world problems this severely limits the number of large clusters that can be feasibly incorporated into the approximation. For example, in the protein design database studied in [18], each node has around 100 states, so even a cluster of only 3 variables would have $10^6$ states. Clearly we cannot use too many such clusters in our approximation.

In this paper we propose a cluster-pursuit method where clusters are incrementally added to the relaxation, and where we only add clusters that are guaranteed to improve the approximation. Similar to the work of [16] who worked on region-pursuit for sum-product generalized belief propagation [19], we show how to use the messages from a given cluster-based approximation to decide which cluster to add next. In addition, by working with a message-passing algorithm based on dual coordinate descent, we monotonically decrease an upper bound on the MAP value.

## 2 MAP and its LP Relaxation

We consider functions over $n$ discrete variables $\boldsymbol{x} = \{x_1, \ldots, x_n\}$ defined as follows. Given a graph $G = (V, E)$ with $n$ vertices, and potentials $\theta_{ij}(x_i, x_j)$ for all edges $ij \in E$, define the function

$$f(\boldsymbol{x}; \boldsymbol{\theta}) = \sum_{ij \in E} \theta_{ij}(x_i, x_j) + \sum_{i \in V} \theta_i(x_i) \ . \tag{1}$$

Our goal is to find the MAP assignment, $\boldsymbol{x}_M$, that maximizes the function $f(\boldsymbol{x}; \boldsymbol{\theta})$.

The MAP problem can be formulated as a linear program as follows. Let $\boldsymbol{\mu}$ be a vector of marginal probabilities that includes $\{\mu_{ij}(x_i, x_j)\}_{ij \in E}$ over variables corresponding to edges and $\{\mu_i(x_i)\}_{i \in V}$ associated with the nodes. The set of $\boldsymbol{\mu}$ that arise from some joint distribution is known as the *marginal polytope* [14],

$$\mathcal{M}(G) = \left\{ \boldsymbol{\mu} \mid \ \exists p(\boldsymbol{x}) \text{ s.t. } \begin{array}{l} p(x_i, x_j) = \mu_{ij}(x_i, x_j) \\ p(x_i) = \mu_i(x_i) \end{array} \right\}.$$

The MAP problem can then be shown to be equivalent to the following LP,

$$\max_{\boldsymbol{x}} f(\boldsymbol{x}, \boldsymbol{\theta}) = \max_{\boldsymbol{\mu} \in \mathcal{M}(G)} \boldsymbol{\mu} \cdot \boldsymbol{\theta} \ , \tag{2}$$

where $\boldsymbol{\mu} \cdot \boldsymbol{\theta} = \sum_{ij \in E} \sum_{x_i, x_j} \theta_{ij}(x_i, x_j) \mu_{ij}(x_i, x_j) + \sum_i \sum_{x_i} \mu_i(x_i) \theta_i(x_i)$. There always exists a maximizing $\boldsymbol{\mu}$ that is integral – a vertex of the marginal polytope – and which corresponds to $\boldsymbol{x}_M$. Although the number of variables in this LP is only $O(|E|+|V|)$, the difficulty comes from an exponential number of linear inequalities typically required to describe the marginal polytope $\mathcal{M}(G)$.

The idea in LP relaxations is to relax the difficult global constraint that the marginals in $\boldsymbol{\mu}$ arise from some common joint distribution. Instead, we enforce this only over some subsets of variables that we refer to as clusters. More precisely, we introduce auxiliary distributions over clusters of variables and constrain the edge distributions $\mu_{ij}(x_i, x_j)$ associated with each cluster to arise as marginals from the cluster distribution.[1] Let $\mathcal{C}$ be a set of clusters such that each $c \in \mathcal{C}$ is a subset of $\{1, \ldots, n\}$, and let $\tau_c(x_c)$ be any distribution over the variables in $c$. We also use $\tau_c(x_i, x_j)$ to refer to the marginal of $\tau_c(x_c)$ for the edge $(i, j)$, i.e. $\tau_c(x_i, x_j) = \sum_{x_{c \setminus i,j}} \tau_c(x_c)$. Define $\mathcal{M}_\mathcal{C}(G)$ as

$$\left\{ \begin{array}{c} \exists \boldsymbol{\tau} \geq 0 \\ \boldsymbol{\mu} \geq 0 \end{array} \middle| \begin{array}{l} \sum_{x_j} \mu_{ij}(x_i, x_j) = \mu_i(x_i) \\ \tau_c(x_i, x_j) = \mu_{ij}(x_i, x_j) \quad \forall c, (i,j) \subseteq c \\ \sum_{x_c} \tau_c(x_c) = 1 \end{array} \right\}$$

It is easy to see that $\mathcal{M}_\mathcal{C}(G)$ is an outer bound on $\mathcal{M}(G)$, namely $\mathcal{M}_\mathcal{C}(G) \supseteq \mathcal{M}(G)$. As we add more clusters to $\mathcal{C}$ the relaxation of the marginal polytope becomes tighter. Note that similar constraints should be imposed on the cluster marginals, i.e., they themselves should arise as marginals from some joint distribution. To exactly represent the marginal polytope, such a hierarchy of auxiliary clusters would require clusters of size equal to the treewidth of the graph. For the purposes of this paper, we will not generate such a hierarchy but instead use the clusters to constrain only the associated edge marginals.

### 2.1 Choosing Clusters in the LP Relaxation

Adding a cluster to the relaxation $\mathcal{M}_\mathcal{C}(G)$ requires computations that scale with the number of possible cluster states. The choice of clusters should therefore be guided by both how much we are able to constrain the marginal polytope, as well as the computational cost of handling larger clusters. We will consider a specific scenario where the clusters are selected from a pre-defined set of possible clusters $\mathcal{C}_0$ such as triplet clusters. However, we will ideally not want to use all of the clusters in $\mathcal{C}_0$, but instead add them gradually based on some ranking criterion.

The best ranking of clusters is problem dependent. In other words, we would like to choose the subset of clusters which will give us the best possible approximation to a particular MAP problem. We seek to *iteratively* improve the approximation, using our current beliefs to guide which clusters to add. The advantage of iteratively selecting the clusters is that we add them only up to the point that the relaxed LP has an integral solution.

Recently, Sontag and Jaakkola [12] suggested an approach for incrementally adding constraints to the marginal polytope using a *cutting-plane algorithm*. A similar approach may in principle be applied to adding clusters to the primal problem. One shortcoming of this approach is that it requires solving the primal LP after every cluster added, and even solving the primal LP once is infeasible for large problems involving hundreds of variables and large state spaces.

In the next section we present a method that incrementally adds clusters, but which works exclusively within the dual LP. The key idea is that the dual LP provides an upper bound on the MAP value, and we seek to choose clusters to most effectively minimize this bound. Note that an analogous bound minimization strategy is problematic in the primal where we would have to assess how much *less* the maximum is due to including additional constraints. In other words, obtaining a certificate for improvement is difficult in the primal. Moreover, unlike the dual, the primal algorithm might not give an upper bound on the MAP prior to convergence.

[1] Each edge may participate in multiple clusters.

Finally, we can "warm start" our optimization scheme after each cluster addition in order to avoid re-solving the dual LP. We do this by reusing the dual variables calculated in the previous iterations which did not have the new clusters.

# 3 Dual LP Relaxation

The obstacles to working in the primal LP lead us to consider the dual of the LP relaxation. Different formulations of the primal LP have lead to different dual LPs, each with efficient message-passing algorithms for solving them [3, 6, 13, 15]. In this paper we focus on a particular dual formulation by Globerson and Jaakkola [3] which has the advantage that the message-passing algorithm corresponds to performing coordinate-descent in the dual LP. Our dual algorithm will address many of the problems that were inherent in the primal approaches, giving us:

1. Monotonically decreasing upper bound on MAP.
2. Choosing clusters which give a guaranteed bound improvement.
3. Simple "warm start" of tighter relaxation.
4. An efficient algorithm that scales to very large problems.

## 3.1 The Generalized MPLP Algorithm

The generalized Max-Product LP (MPLP) message-passing algorithm, introduced in [3], decreases the dual objective of the cluster-based LP relaxation at every iteration. This monotone property makes it ideal for adding clusters since we can initialize the new messages such that the dual value is monotonically decreased.

Another key advantage of working in the dual is that the dual objective gives us a certificate of optimality. Namely, if we find an assignment $\boldsymbol{x}$ such that $f(\boldsymbol{x};\boldsymbol{\theta})$ is equal to the dual objective, we are guaranteed that $\boldsymbol{x}$ is the MAP assignment (since the dual objective upper bounds the MAP value). Indeed, using this property we show in our experiments that MAP assignments can be found for nearly all of the problems we consider.

We next describe the generalized MPLP algorithm for the special case of clusters comprised of three nodes. Although the algorithm applies to general clusters, we focus on triplets for simplicity, and because these are the clusters used in the current paper.

MPLP passes the following types of messages:

- **Edge to Node**: For every edge $e \in E$ ($e$ denotes two indices in $V$) and every node $i \in e$, we have a message $\lambda_{e\to i}(x_i)$.
- **Edge to Edge**: For every edge $e \in E$, we have a message $\lambda_{e\to e}(x_e)$ (where $x_e$ is shorthand for $x_i, x_j$, and $i$ and $j$ are the nodes in the edge).
- **Triplet to Edge**: For every triplet cluster $c \in \mathcal{C}$, and every edge $e \in c$, we have a message $\lambda_{c\to e}(x_e)$.

The updates for these messages are given in Figure 1. To guarantee that the dual objective decreases, all messages from a given edge must be sent simultaneously, as well as all messages from a triplet to its three edges.

The dual objective that is decreased in every iteration is given by

$$\begin{aligned} g(\boldsymbol{\lambda}) &= \sum_{i\in V} \max_{x_i} \left[\theta_i(x_i) + \sum_{k\in N(i)} \lambda_{ki\to i}(x_i)\right] \\ &+ \sum_{e\in E} \max_{x_e} \left[\lambda_{e\to e}(x_e) + \sum_{c:e\in c} \lambda_{c\to e}(x_e)\right] \end{aligned}$$

It should be noted, however, that not all $\boldsymbol{\lambda}$ are dual feasible. Rather, $\boldsymbol{\lambda}$ needs to result from a reparameterization of the underlying potentials (see [3]). However, it turns out that after updating all the MPLP messages once, all subsequent $\boldsymbol{\lambda}$ will be dual feasible, regardless of how $\boldsymbol{\lambda}$ is initialized.[2]

By LP duality, there exists a value of $\boldsymbol{\lambda}$ such that $g(\boldsymbol{\lambda})$ is equal to the optimum of the corresponding primal LP. Although the MPLP updates decrease the objective at every iteration, they may converge to a $\boldsymbol{\lambda}$ that is not dual optimal, as discussed in [3]. However, as we will show in the experiments, our procedure often finds the exact MAP solution, and therefore also achieves the primal optimum in these cases.

## 3.2 Choosing Clusters in the Dual LP Relaxation

In this section we provide a very simple procedure that allows adding clusters to MPLP, while satisfying the

[2] In our experiments, we initialize all messages to zero.

- **Edge to Node**: For every edge $ij \in E$ and node $i$ (or $j$) in the edge:

$$\lambda_{ij\to i}(x_i) \leftarrow -\frac{2}{3}\Big(\lambda_i^{-j}(x_i) + \theta_i(x_i)\Big) + \frac{1}{3}\max_{x_j}\Big[\sum_{c:ij\in c}\lambda_{c\to ij}(x_i,x_j) + \lambda_j^{-i}(x_j) + \theta_{ij}(x_i,x_j) + \theta_j(x_j)\Big]$$

where $\lambda_i^{-j}(x_i)$ is the sum of edge-to-node messages into $i$ that are not from edge $ij$, namely: $\lambda_i^{-j}(x_i) = \sum_{k\in N(i)\setminus j}\lambda_{ik\to i}(x_i)$.

- **Edge to Edge**: For every edge $ij \in E$:

$$\lambda_{ij\to ij}(x_i,x_j) \leftarrow -\frac{2}{3}\sum_{c:ij\in c}\lambda_{c\to ij}(x_i,x_j) + \frac{1}{3}\Big[\lambda_j^{-i}(x_j) + \lambda_i^{-j}(x_i) + \theta_{ij}(x_i,x_j) + \theta_i(x_i) + \theta_j(x_j)\Big]$$

- **Triplet to Edge**: For every triplet $c \in \mathcal{C}$ and every edge $e \in c$:

$$\lambda_{c\to e}(x_e) \quad\leftarrow\quad -\frac{2}{3}\Big(\lambda_{e\to e}(x_e) + \sum_{\substack{c' \neq c \\ e \in c'}}\lambda_{c'\to e}(x_e)\Big) + \frac{1}{3}\max_{x_{c\setminus e}}\Big[\sum_{e'\in c\setminus e}\Big(\lambda_{e'\to e'}(x_{e'}) + \sum_{\substack{c' \neq c \\ e' \in c'}}\lambda_{c'\to e'}(x_{e'})\Big)\Big]$$

Figure 1: The generalized MPLP updates for an LP relaxation with three node clusters.

algorithmic properties in the beginning of Section 3.

Assume we have a set of triplet clusters $\mathcal{C}$ and now wish to add a new triplet. Denote the messages before adding the new triplet by $\boldsymbol{\lambda}_t$. Two questions naturally arise. The first is: assuming we decide to add a given triplet, how do we set $\boldsymbol{\lambda}_{t+1}$ such that the dual objective retains its previous value $g(\boldsymbol{\lambda}_t)$. The second question is how to choose the new triplet to add.

The initialization problem is straightforward. Simply set $\boldsymbol{\lambda}_{t+1}$ to equal $\boldsymbol{\lambda}_t$ for all messages from triplets and edges in the previous run, and set $\boldsymbol{\lambda}_{t+1}$ for the messages from the new triplet to its edges to zero.[3] This clearly results in $g(\boldsymbol{\lambda}_{t+1}) = g(\boldsymbol{\lambda}_t)$.

In order to choose a *good* triplet, one strategy would be to add different triplets and run MPLP until convergence to find the one that decreases the objective the most. However, this may be computationally costly and, as we show in the experiments, is not necessary. Instead, the criterion we use is to consider the decrease in value that results from just sending messages from the triplet $c$ to its edges (while keeping all other messages fixed).

The decrease in $g(\boldsymbol{\lambda})$ resulting from such an update has a simple form, as we show next. Assume we are considering adding a triplet $c$. For every edge $e \in c$, define $b_e(x_e)$ to be

$$b_e(x_e) = \lambda_{e\to e}(x_e) + \sum_{c':e\in c'}\lambda_{c'\to e}(x_e)\ , \qquad (3)$$

[3]It is straightforward to show that $\boldsymbol{\lambda}_{t+1}$ is dual feasible.

where the summation over clusters $c'$ does not include $c$ (those messages are initially zero). The decrease in $g(\boldsymbol{\lambda})$ corresponding to updating only messages from $c$ to the edges $e \in c$ can be shown to be

$$d(c) = \sum_{e\in c}\max_{x_e} b_e(x_e) - \max_{x_c}\left[\sum_{e\in c} b_e(x_e)\right]\ . \qquad (4)$$

The above corresponds to the difference between independently maximizing each edge and jointly maximizing over the three edges. Thus $d(c)$ is a lower bound on the improvement in the dual objective if we were to add triplet $c$. Our algorithm will therefore add the triplet $c$ that maximizes $d(c)$.

### 3.3 The Dual Algorithm

We now present the complete algorithm for adding clusters and optimizing over them. Let $\mathcal{C}_0$ be the predefined set of triplet clusters that we will consider adding to our relaxation, and let $\mathcal{C}_L$ be the initial relaxation consisting of only edge clusters (pairwise local consistency).

**1.** Run MPLP until convergence using the $\mathcal{C}_L$ clusters.

**2.** Find an integral solution $\boldsymbol{x}$ by locally maximizing the single node beliefs $b_i(x_i)$, where $b_i(x_i) = \theta_i(x_i) + \sum_{k\in N(i)}\lambda_{ki\to i}(x_i)$. Ties are broken arbitrarily.

**3.** If the dual objective $g(\boldsymbol{\lambda}_t)$ is sufficiently close to the primal objective $f(\boldsymbol{x};\theta)$, terminate (since $\boldsymbol{x}$ is approximately the MAP).

**4.** Add the cluster $c \in \mathcal{C}_0$ with the largest guaranteed bound improvement, $d(c)$, to the relaxation.

**5.** Construct "warm start" messages $\boldsymbol{\lambda}_{t+1}$ from $\boldsymbol{\lambda}_t$.

**6.** Run MPLP for $N$ iterations, and return to **2**.

Note that we obtain (at least) the promised bound improvement $d(c)$ within the first iteration of step **6**. By allowing MPLP to run for $N$ iterations, the effect of adding the cluster will be propagated throughout the model, obtaining an additional decrease in the bound. Since the MPLP updates correspond to coordinate-descent in the dual LP, every step of the algorithm decreases the upper bound on the MAP. The monotonicity property holds even if MPLP does not converge in step **6**, giving us the flexibility to choose the number of iterations $N$. In Section 5 we show results corresponding to two different choices of $N$.

In the case where we run MPLP to convergence before choosing the next cluster, we can show that the greedy bound minimization corresponds to a cutting-plane algorithm, as stated below.

**Theorem 1.** *Given a dual optimal solution, if we find a cluster for which we can guarantee a bound decrease, all primal optimal solutions were inconsistent with respect to this cluster.*

*Proof.* By duality both the dual optimum and the primal optimum will decrease. Suppose for contradiction that in the previous iteration there was a primal feasible point that was cluster consistent and achieved the LP optimum. Since we are maximizing the LP, after adding the cluster consistency constraint, this point is still feasible and the optimal value of the primal LP will not change, giving our contradiction. □

This theorem does not tell us *how much* the given cluster consistency constraint was violated, and the distinction remains that a typical cutting-plane algorithm would attempt to find the constraint which is most violated.

## 4 Related Work

Since MPLP is closely related to the max-product generalized belief propagation (GBP) algorithm, our work can be thought of as a region-pursuit method for GBP. This is closely related to the work of Welling [16] who suggested a region-pursuit method for sum-product GBP. Similar to our work, he suggested greedily adding from a candidate set of possible clusters. At each iteration, the cluster that results in the largest change in the GBP free energy is added. He showed excellent results for 2D grids, but on fully connected graphs the performance actually started deteriorating with additional clusters. In [17], a heuristic related to maxent normality [19] was used as a stopping criterion for region-pursuit to avoid this behavior. In our work, in contrast, since we are working with the dual function of the LP, we can guarantee monotonic improvement throughout the running of the algorithm.

Our work is also similar to Welling's in that we focus on criteria for determining the utility of adding a cluster, not on *finding* these clusters efficiently. We found in our experiments that a simple enumeration over small clusters proved extremely effective. For problems where triplet clusters alone would not suffice to find the MAP, we could triangulate the graph and consider larger clusters. This approach is reminiscent of the bounded join-graphs described in [1].

There is a large body of recent work describing the relationship between message-passing algorithms such as belief propagation, and LP relaxations [7, 15, 18]. Although we have focused here on using one particular message-passing algorithm, MPLP, we emphasize that similar region-pursuit algorithms can be derived for other message-passing algorithms as well. In particular, for all the convex max-product BP algorithms described in [15], it is easy to design region-pursuit methods. The main advantage of using MPLP is its guaranteed decrease of the dual value at each iteration, a guarantee that does not exist for general convex BP algorithms.

Region-pursuit algorithms are also conceptually related to the question of message scheduling in BP, as in the work of Elidan et al. [2]. One way to think of region-pursuit is to consider a graph where all the clusters are present all the time, but send and receive non-informative messages. The question of which cluster to add to an approximation, is thus analogous to the question of which message to update next.

## 5 Experiments

Due to the scalable nature of our message-passing algorithm, we can apply it to cases where standard LP solvers cannot be applied to the primal LP (see also [18]). Here we report applications to problems in computational biology and machine vision.[4]

We use the algorithm from Section 3.3 for all of our experiments. We first run MPLP with edge clusters until convergence or for at most 1000 iterations, whichever comes first. All of our experiments, except those intended to show the difference between schedules, use $N = 20$ for the number of MPLP iterations run after adding a cluster. While running MPLP we use the messages to decode an integral solution, and compare the dual objective to the value of the integral solution. If these are equal, we have found the MAP solution.[5] Otherwise, we keep adding triplets.

[4]Graphical models for these are given in [18].

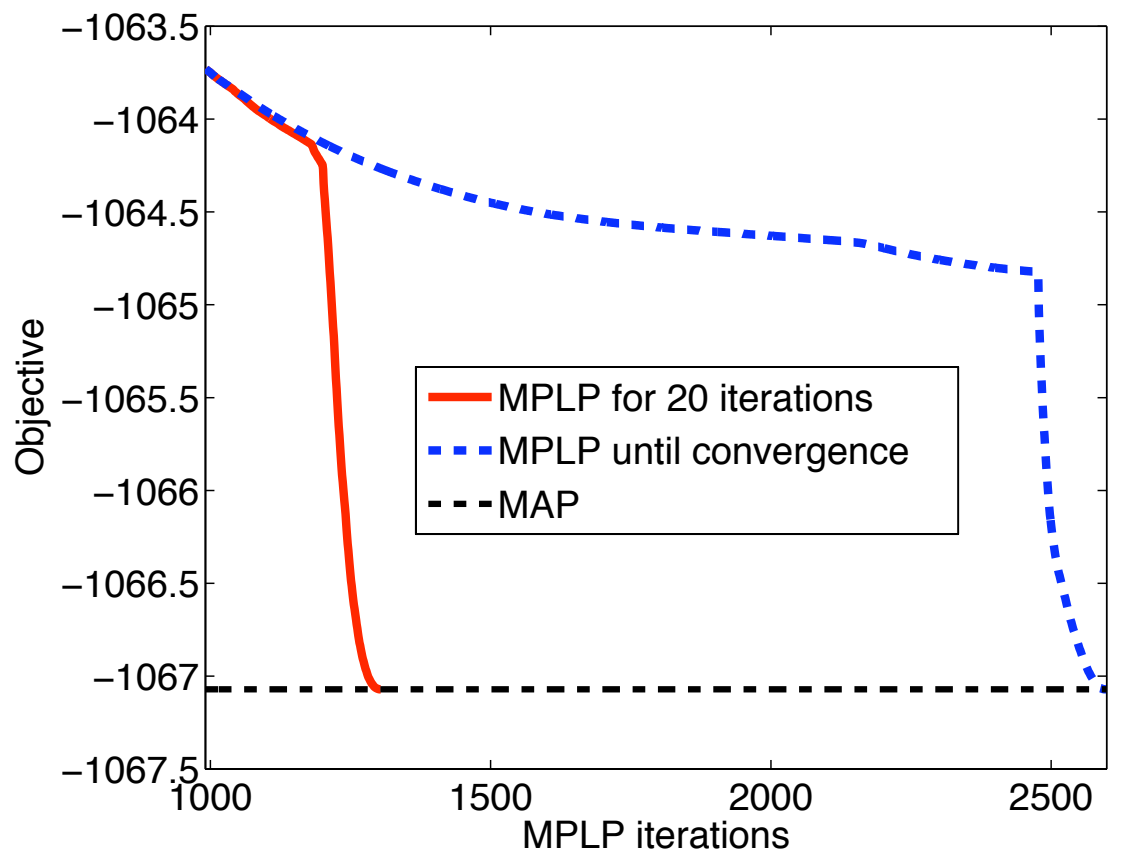

Figure 2: Comparison of different schedules for adding clusters to relaxation on a side-chain prediction problem.

Our results will show that we often find the MAP solution to these hard problems by using only a small number of triplet clusters. This indicates both that triplets are sufficient for characterizing $\mathcal{M}(G)$ near the MAP solution of these problems, and that our algorithm can efficiently find the informative triplets.

### 5.1 Side-Chain Prediction

The side-chain prediction problem involves finding the three-dimensional configuration of rotamers given the backbone structure of a protein [18]. This problem can be posed as finding the MAP configuration of a pairwise model, and in [18] the TRBP algorithm [13] was used to find the MAP solution for most of the models studied. However, for 30 of the models, TRBP could not find the MAP solution.

In earlier work [12] we used a cutting-plane algorithm to solve these side-chain problems and found the MAP solution for all 30 models. Here, we applied our dual algorithm to the same 30 models and found that it also results in the MAP solution for all of them (up to a $10^{-4}$ integrality gap). This required adding between 1 and 27 triplets per model. The running time was between 1 minute and 1 hour to solve each problem, with over half solved in under 9 minutes. On average we added only 7 triplets (median was 4.5), another indication of the relative ease with which these techniques can solve the side-chain prediction problem.

[5]In practice, we terminate when the dual objective is within $10^{-4}$ of the decoded assignment, so these are approximate MAP solutions. Note that the objective values are significantly larger than this threshold.

We also used these models to study different update schedules. One schedule (which gave the results in the previous paragraph) was to first run a pairwise model for 1000 iterations, and then alternate between adding triplets and running MPLP for 20 more iterations. In the second schedule, we run MPLP to convergence after adding each triplet. Figure 2 shows the two schedules for the side-chain protein '1gsk', one of the side-chain proteins which took us the longest to solve (30 minutes). Running MPLP to convergence results in a much larger number of overall MPLP iterations compared to using only 20 iterations. This highlights one of the advantages of our method: adding a new cluster does not require solving the earlier problem to convergence.

### 5.2 Protein Design

The protein design problem is the inverse of the protein folding problem. Given a particular 3D shape, we wish to find a sequence of amino-acids that will be as stable as possible in that 3D shape. Typically this is done by finding a set of amino-acids and rotamer configurations that minimizes an approximate energy.

While the problem is quite different from side-chain prediction, it can be solved using the same graph structure, as shown in [18]. The only difference is that now the nodes do not just denote rotamers, but also the identity of the amino-acid at that location. Thus, the state-space here is significantly larger than in the side-chain prediction problem (up to 180 states per variable for most variables).

In contrast to the side-chain prediction problems, which are often easily solved by general purpose integer linear programming packages such as CPLEX's branch-and-cut algorithm [5], the sheer size of the protein design problems immediately limits the techniques by which we can attempt to solve them. Algorithms such as our earlier cutting-plane algorithm [12] or CPLEX's branch-and-cut algorithm require solving the primal LP relaxation at least once, but solving the primal LP on all but the smallest of the design problems is intractable [18]. Branch and bound schemes have been recently used in conjunction with a message passing algorithm [4] and applied to similar protein design problems, although not the ones we solve here.

We applied our method to the 97 protein design problems described in [18], adding 5 triplets at a time to the relaxation. The key striking result of these experiments is that our method found the exact MAP configuration for all but one of the proteins[6] (up to a precision of $10^{-4}$ in the integrality gap). This is es-pecially impressive since, as reported in [18], only 2 of these problems were solvable using TRBP, and the primal problem was too big for commercial LP solvers such as CPLEX. For the problem where we did not find the MAP, we did not reach a point where all the triplets in the graph were included, since we ran out of memory beforehand.

[6]We could not solve '1fpo', the largest protein.

Among the problems that were solved exactly, the mean running time was 9.7 hours with a maximum of 11 days and a minimum of a few minutes. We note again that most of these problems could not be solved using LP solvers, and when LP solvers could be used, they were typically at least 10 times slower than message-passing algorithms similar to ours (see [18] for detailed timing comparisons).

Note that the main computational burden in the algorithm is processing triplet messages. Since each variable has roughly 100 states, passing a triplet message requires $10^6$ operations. Thus the number of triplets added is the key algorithmic complexity issue. For the models that were solved exactly, the median number of triplets added was 145 (min: 5, max: 735). As mentioned earlier, for the unsolved model this number grew until the machine's memory was exhausted. We believe however, that by optimizing our code for speed and memory we will be able to accommodate a larger number of triplets, and possibly solve the remaining model as well. Our current code is written mostly in Matlab, so significant optimization may be possible.

### 5.3 Stereo Vision

Given a stereo pair of images, the stereo problem is to find the disparity of each pixel in a reference image. This disparity can be straightforwardly translated into depth from the camera. The best algorithms currently known for the stereo problem are those that minimize a global energy function [10], which is equivalent to finding a MAP configuration in a pairwise model.

For our experiments we use the pairwise model described in [18], and apply our procedure to the "Tsukuba" sequence from the standard Middlebury stereo benchmark set [10], reduced in size to contain 116x154 pixels.

Since there are no connected triplets in the grid graph, we use our method with square clusters. We calculate the bound decrease using square clusters, but rather than add them directly, we triangulate the cycle and add two triplet clusters. This results in an equivalent relaxation, but has the consequence that we may have to wait until MPLP convergence to achieve the guaranteed bound improvement.

In the first experiment, we varied the parameters of the energy function to create several different instances. We tried to find the MAP using TRBP, resolving ties using the methods proposed in [8]. In 4 out of 10 cases those methods failed. Using our algorithm, we managed to find the MAP for all 4 cases.[7]

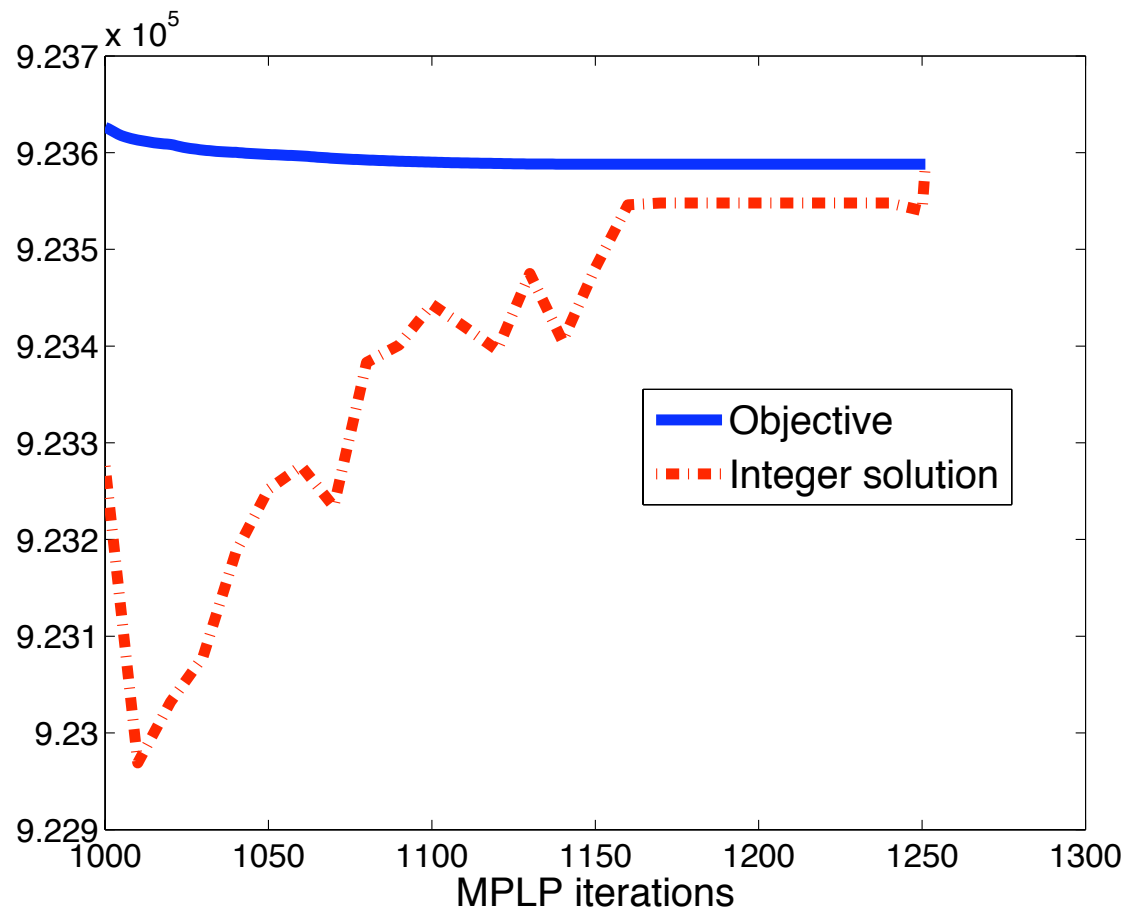

Figure 3: Dual objective and value of decoded integer solution for one of the reduced "Tsukuba" stereo models, as a function of MPLP iterations. It can be seen that both curves converge to the same value, indicating that the MAP solution was found.

Figure 3 shows the dual objective and the decoded integer solution after each MPLP iteration, for one set of parameters.

In the results above, we added 20 squares at a time to the relaxation. We next contrasted it with two strategies: one where we pick 20 random squares (not using our bound improvement criterion) and one where we pick the single best square according to the bound criterion. Figure 4 shows the resulting bound per iteration for one of the models. It can be seen that the random method is much slower than the bound criterion based one, and that adding 20 squares at a time is better than just one. We ended up adding 1060 squares when adding 20 at a time, and 83 squares when adding just one. Overall, adding 20 squares at a time turned out to be faster.

We also tried running MPLP with all of the square clusters. Although fewer MPLP iterations were needed, the cost of using all squares resulted in an overall running time of about four times longer.

[7] For one of these models, a few single node beliefs at convergence were tied, and we used the junction tree algorithm to decode the tied nodes (see [8]).

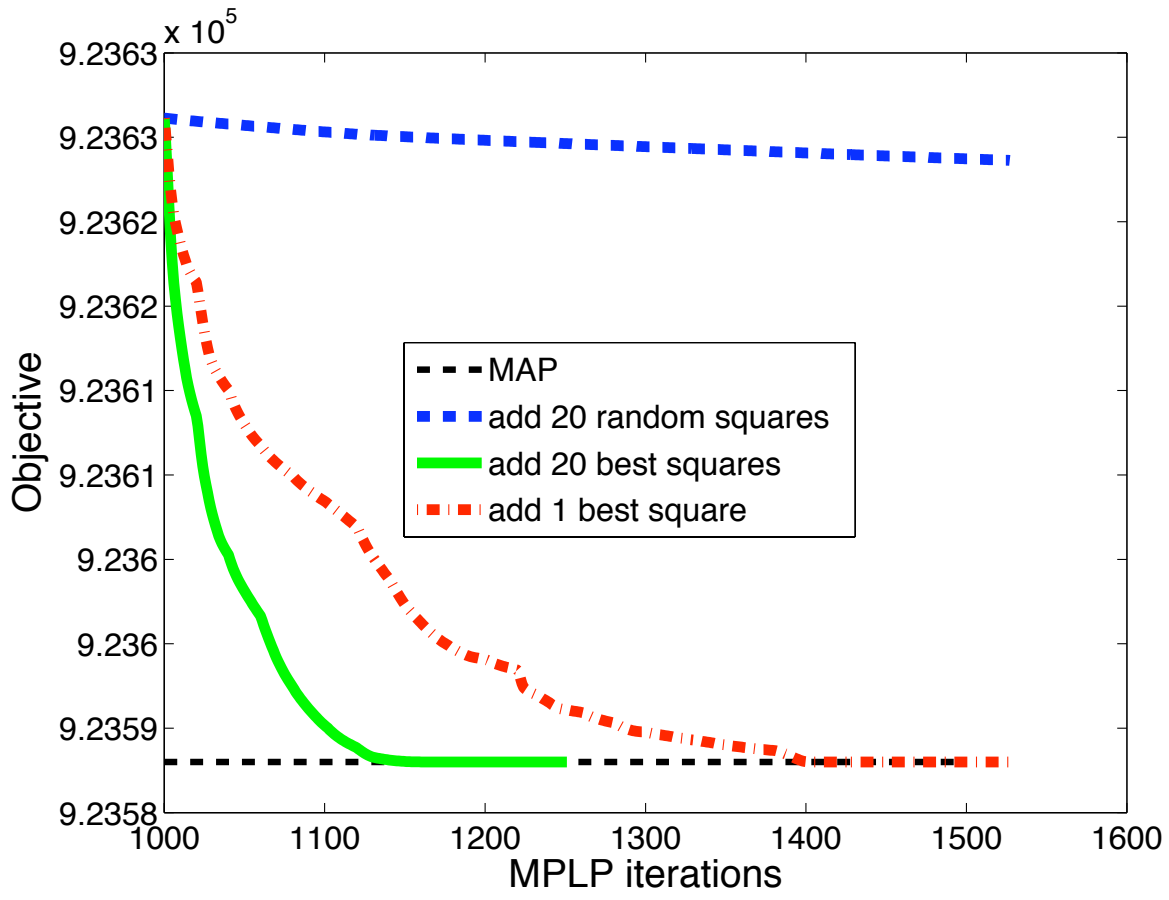

Figure 4: Comparison of different schedules for adding squares in one of the stereo problems.

## 6 Conclusion

In order to apply LP relaxations to real-world problems, one needs to find an efficient way of adding clusters to the basic relaxation such that the problem remains tractable but yields a better approximation of the MAP value.

In this paper we presented a greedy bound-minimization algorithm on the dual LP to solve this problem, and showed that it has all the necessary ingredients: an efficient message-passing algorithm, "warm start" of the next iteration using current beliefs, and a monotonically decreasing bound on the MAP.

We showed that the algorithm works well in practice, finding the MAP configurations for many real-world problems that were previously thought to be too difficult for known methods to solve. While in this paper we focused primarily on adding triplet clusters, our approach is general and can be used to add larger clusters as well, as long as as the messages in the dual algorithm can be efficiently computed.

Finally, while here we focused on the MAP problem, similar ideas may be applied to approximating the marginals in graphical models.

### Acknowledgements

This work was supported in part by the DARPA Transfer Learning program and by the Israeli Science Foundation. D.S. was also supported by a NSF Graduate Research Fellowship.